\documentclass[conference]{IEEEtran}
\IEEEoverridecommandlockouts
\usepackage{cite}
\usepackage{xurl}
\usepackage{multirow}
\usepackage{url} 
\usepackage{amsmath,amssymb,amsfonts}
\usepackage{algorithmic}
\usepackage{graphicx}
\usepackage{textcomp}
\usepackage{xcolor}
\usepackage{mathrsfs}
\usepackage{amsthm}  
\usepackage{booktabs}
\usepackage[hidelinks]{hyperref}

\usepackage[final]{changes}

\newtheorem{property}{Property}
\def\BibTeX{{\rm B\kern-.05em{\sc i\kern-.025em b}\kern-.08em
    T\kern-.1667em\lower.7ex\hbox{E}\kern-.125emX}}
\begin{document}

\title{Curtail Renewables to Enhance Flexibility: \\ A \added{Regulated Forecast-based} Dispatch Approach

\thanks{The work was supported in part by the National Natural Science Foundation of China under Grant 52377105. (Corresponding author: Ye Guo, e-mail: ye.guo@polyu.edu.hk)}
}

\author{
\IEEEauthorblockN{Zhiyi Zhao\textsuperscript{1}, Ye Guo\textsuperscript{2}, Zhenjia Lin\textsuperscript{2}, Yinliang Xu\textsuperscript{1}}
\IEEEauthorblockA{\textsuperscript{1}\textit{Shenzhen International Graduate School, Tsinghua University, Shenzhen, China} \\
\textsuperscript{2}\textit{Department of Building Environment and Energy Engineering, Hong Kong Polytechnic University, Kowloon, Hong Kong} \\
Emails: zhiyi-zh24@mails.tsinghua.edu.cn, ye.guo@polyu.edu.hk, epjack.lin@polyu.edu.hk, xu.yinliang@sz.tsinghua.edu.cn}
}

\maketitle

\begin{abstract} 
\added{This paper considers the flexibility degradation problem caused by excessive flexible ramping product (FRP) requirements with high variable energy resource (VER) penetration}. Based on the rolling-window co-optimization model of energy and FRP, theoretical analysis of this paper reveals a unit dispatch transfer effect, in which high FRP requirements under forecast-based dispatch (FBD) constrain real-time flexibility and distort economic efficiency. To alleviate this effect, a regulated forecast-based dispatch (RFBD) approach is proposed, which moderately caps VER outputs and enhances system flexibility. Simulation results demonstrate that the proposed approach effectively lowers FRP requirements and reduces operating cost compared with FBD. 
\end{abstract}


\section{Introduction}

Real-time electricity markets commonly adopt a multi-interval rolling-window dispatch framework, as implemented by CAISO \cite{CAISO_BPMO_V102_Redline} and NYISO \cite{NYISO_LRR}. In this framework, the scheduling horizon is divided into a \emph{binding interval}, whose dispatch is executed in real time, and several \emph{advisory intervals}, which provide non-binding guidance for future scheduling \cite{YG}.

In conventional rolling-window dispatch, independent system operators (ISOs) apply a \emph{forecast-based dispatch} (FBD) mode, where variable energy resources (VERs) such as wind and solar power are scheduled according to forecasted outputs \cite{SHI2024122435}. Forecasts for the binding interval are typically assumed accurate, while those for advisory intervals are uncertain. To manage such uncertainty, CAISO introduced the \emph{Flexible Ramping Product} (FRP), including flexible ramping up (FRU) and flexible ramping down (FRD) \cite{CAISO_FRP_Deliverability}. FRP reserves ramping capability to accommodate forecast deviations when advisory intervals roll into binding intervals, with requirements determined from historical forecast errors.

Previous studies have investigated FRP procurement frameworks  \cite{DayAhead}, bidding strategies for flexible resources  \cite{EV}, and multi-timescale requirement estimation methods \cite{CUI201827}. However, the impact of FRP requirements on system flexibility and economic efficiency remains insufficiently explored. As VER penetration increases, forecast uncertainty amplifies, leading to higher FRP requirements in real-time dispatch.

To fill this research gap, this paper formulates a rolling-window co-optimization model for energy and FRP under the FBD framework. Theoretical analysis of the dispatch optimality identifies the \emph{unit dispatch transfer} effect, where high FRP requirements limit real-time flexibility and result in underutilization of cost-efficient units. To mitigate this effect, a \emph{regulated forecast-based dispatch}  (RFBD) approach is proposed. This approach offers two key benefits, forming the main contributions of the paper: 1) it reduces uncertainty, as reflected in lower FRP requirements; 2) it enhances system flexibility by relaxing FRP constraints on cost-efficient units and facilitating \emph{cross-feeding} among VER units, allowing renewables to partially offset each other’s variability in binding intervals and ultimately reduce operating cost.

\section{Model Formulation}

\subsection{Rolling-Window Dispatch Framework}

\begin{figure}[ht]
\centering
\includegraphics[width=0.5\textwidth]{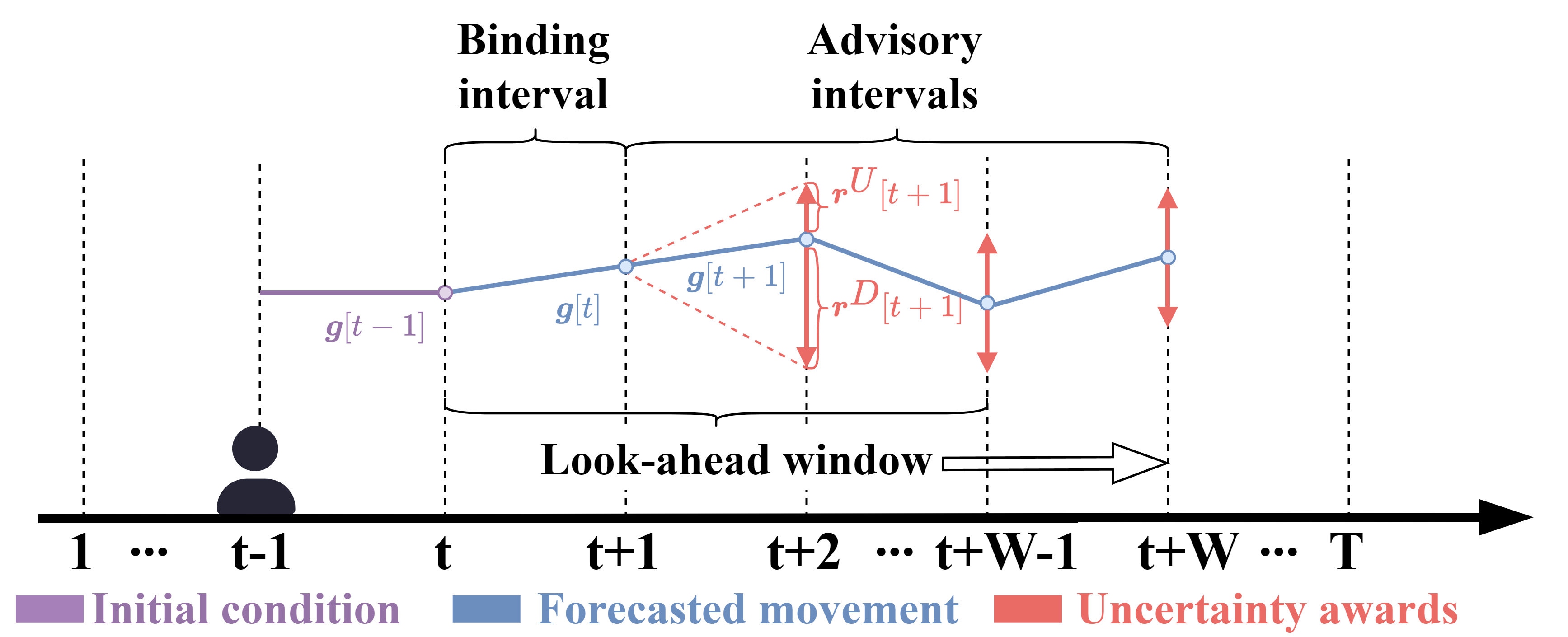}
\caption{Rolling-window dispatch framework}
\label{fig:rolling}
\end{figure}

In the real-time market with a rolling optimization framework, the scheduling horizon $\mathscr{H} = \{1, \dots, T\}$ is divided into unit-length intervals, each representing $[t, t+1)$. To incorporate continuously updated forecasts, the system operator clears the market through sequential look-ahead co-optimizations over sliding windows of $W$ intervals, as illustrated in Fig.~\ref{fig:rolling}.

At time $t-1$, a joint energy, ancillary services and FRP co-optimization is performed over the window $\mathscr{H}_t = \{t, \dots, t+W-1\}$, consistent with CAISO’s market design. The first interval ($\tau = t$) is binding and implemented in real time, while $\tau = t+1, \dots, t+W-1$ are advisory and non-binding. When the remaining horizon is shorter than $W$, the window is truncated accordingly. As time advances, the ISO executes the binding results for each interval and updates forecasts for the next window $\mathscr{H}_{t+1} = \{t+1, \dots, t+W\}$.

\subsection{Energy-FRP Co-optimization Model}
Building on the rolling-window framework introduced above, we formulate a multi-interval co-optimization model for energy and FRP under the following assumptions: (1) generator energy bid-in cost functions are linear; (2) VERs are modeled as negative loads; (3) transmission congestion is neglected; and (4)  ancillary services such as regulation and spinning reserves are not considered.

The co-optimization model for \(\mathscr{H}_{t}\) can be formulated as \cite{SHI2024122435, CAISO_FRP_Deliverability}:

\begin{alignat}{2} 
&\text{minimize } F^t(g_{i\tau}, \delta d_{l\tau}) = \nonumber \\ 
&\sum_{\tau \in \mathscr{H}_t} \sum_{i \in N_G} C_{i\tau}^g g_{i\tau} 
+ \sum_{\tau \in \mathscr{H}_i} \sum_{l \in N_D} C_{l\tau}^L \delta d_{l\tau}, 
\label{eq:obj} \\ 
&\text{subject to } \forall \tau \in \mathscr{H}_t: \nonumber \\ 
&\text{Power balance constraints:} \nonumber \\ 
&\lambda_{\tau}: \sum_{i \in N_G} g_{i\tau} 
= \sum_{l \in N_D} (\hat{d}_{l\tau|t} -\delta d_{l\tau})-  \sum_{v \in \{w,s\}} \sum_{k \in N_v} \hat{v}_{k\tau|t},
\label{eq:balance} \\
&\text{Flexible ramping requirements:} \nonumber \\ 
&\phi_{\tau}^U: \sum_{i \in N_G} r_{i\tau}^U = R_{\tau}^U, 
 \label{eq:fru}\\ 
&\phi_{\tau}^D: \sum_{i \in N_G} r_{i\tau}^D = R_{\tau}^D, 
 \label{eq:frd}\\
&\text{Generation limits } \forall i \in N_{G}: \nonumber \\ 
&(\underline{\nu}_{i\tau}, \overline{\nu}_{i\tau}) : 
\underline{G}_{i\tau} \leq g_{i\tau} - r^{D}_{i\tau},\;
g_{i\tau} + r^{U}_{i\tau} \leq \overline{G}_{i\tau}, 
\label{eq:gen_limit}\\ 
&\text{Ramping capability limit } \forall i \in N_{G}: \nonumber \\ 
&\underline{\rho}_{i\tau } : -\underline{r}_{i\tau} 
\leq g_{i\tau} - g_{i(\tau - 1)} - r_{i\tau}^D , 
\label{eq:ramp_down}\\ 
&\overline{\rho}_{i\tau} : g_{i\tau} - g_{i(\tau - 1)} + r_{i\tau}^U 
\leq \overline{r}_{i\tau}, \label{eq:ramp_up} \\
&\text{Load shedding constraint } \forall l \in N_{D}: \nonumber \\ 
&(\underline{\gamma}_{l\tau},\overline{\gamma}_{l\tau}) : 0
\leq \delta d_{l\tau} \leq  \hat{d}_{l\tau} .
\label{eq:load_shed}
\end{alignat}

The objective function \eqref{eq:obj} minimizes the total generation cost and load-shedding penalty. The constraints are summarized as follows: 
Equation~\eqref{eq:balance} ensures supply–demand equilibrium, where $\hat{d}_{l\tau|t}$ and $\hat{v}_{k\tau|t}$ denote the ISO’s load and VER forecasts of interval $\tau$ standing at time $t-1$, and VERs (wind and solar) are modeled as negative loads.
Equations~\eqref{eq:fru}--\eqref{eq:frd} guarantee sufficient upward ($R_\tau^U$) and downward ($R_\tau^D$) ramping capability. For the binding interval ($\tau = t$), $R_t^U = R_t^D = 0$, while for advisory intervals ($\tau = t+1,\dots,t+W-1$), requirements are derived from historical forecast errors (see Section~\ref{sec:FRP_req}). 
Equation~\eqref{eq:gen_limit} restricts scheduled generation and FRU/FRD allocations within capacity bounds, and 
Equations~\eqref{eq:ramp_down}--\eqref{eq:ramp_up} maintain inter-temporal feasibility by constraining generation changes between consecutive intervals within ramping limits.

\subsection{FRP Requirements Determination} \label{sec:FRP_req}
The FRP requirements $R_{\tau}^{U}$ and $R_{\tau}^{D}$ in the co-optimization model are determined by the ISO based on historical forecast errors of load and VERs. In CAISO, FRP requirements are obtained using a Mosaic Quantile Regression (MQR) method, which quantifies uncertainty from demand, solar, wind, and net load forecast errors, updated daily with rolling datasets \cite{CAISO_Appendices_Market_Operations_V62}.

\begin{figure}[ht]
\centering
\includegraphics[width=0.4\textwidth]{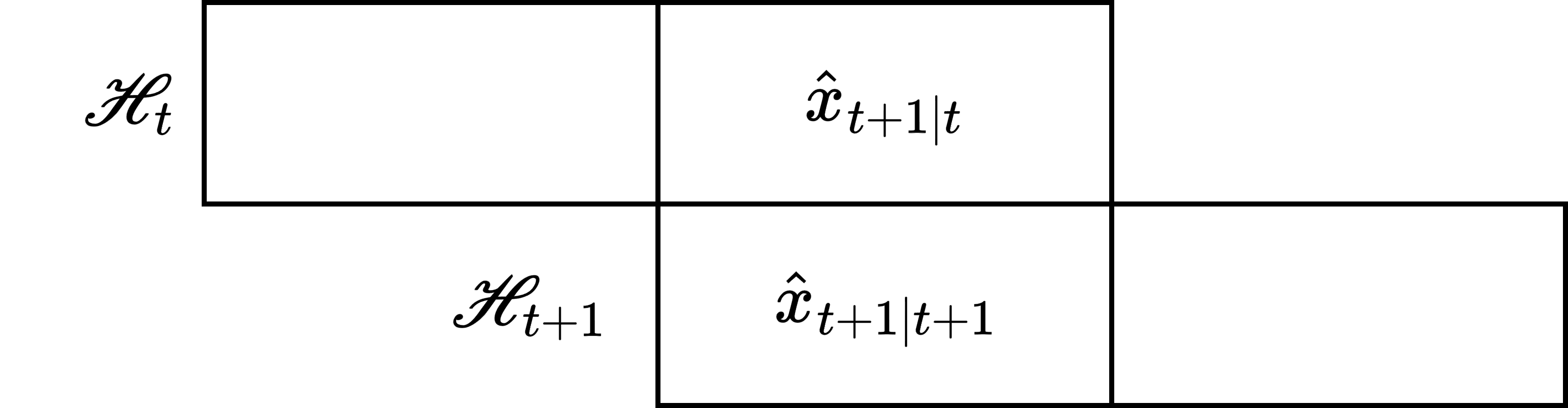}
\caption{Construction of 5-min forecast error.}
\label{fig:uncertainty_measure}
\end{figure}

For a look-ahead window $\mathscr{H}_t = \{t, \dots, t+W-1\}$, the forecast for the first advisory interval $t+1$ is denoted by $\hat{x}_{t+1|t}$ and updated to $\hat{x}_{t+1|t+1}$ when interval $t+1$ becomes binding, as shown in Fig.~\ref{fig:uncertainty_measure}. The net load forecast error for interval $t+1$  is thus expressed as
\begin{equation}
\Delta N_{t+1} = \sum_{l \in N_D} \big( \hat{d}_{l,t+1|t+1} - \hat{d}_{l,t+1|t} \big)
                 - \sum_{k \in N_v} \big( \hat{v}_{k,t+1|t+1} - \hat{v}_{k,t+1|t} \big),
\end{equation}

Based on the upper and lower quantiles of net load forecast error distribution across historical samples, the FRP requirements are then set using MQR method.

\subsection{Unit Dispatch Transfer}\label{sec:Unit_Dispatch_Transfer}

There is no bidding of FRP; its prices are set as the  marginal cost 
\(\phi_{\tau}^U = \frac{\partial L}{\partial R_\tau^U}, \quad
\phi_{\tau}^D = \frac{\partial L}{\partial R_\tau^D}\) \cite{CAISO_BPMO_V102_Redline},
quantifying the marginal impact of increasing FRU/FRD requirements on the system's optimal cost.

The KKT stationarity conditions for FRP variables are:
\begin{align}
\frac{\partial L}{\partial r_{i\tau}^U} 
&= -\phi_{\tau}^U + \overline{\nu}_{i\tau} + \overline{\rho}_{i\tau} = 0, \quad
\frac{\partial L}{\partial r_{i\tau}^D} 
= -\phi_{\tau}^D + \underline{\nu}_{i\tau} + \underline{\rho}_{i\tau} = 0.
\label{eq:KKT_FRP_compact}
\end{align}

Treating all conventional generators as an aggregate FRP provider, accommodating additional FRP without triggering load shedding requires non-binding aggregate ramping and capacity margins. Combined with \eqref{eq:KKT_FRP_compact}, the minimal feasible condition for a non-zero $\phi_{\tau}^U$ (analogously $\phi_{\tau}^D$) occurs when at least two generators are binding on different limits:
\begin{equation}
\exists i,j \in N_G,\, i\neq j:\;
\overline{\nu}_{i\tau} > 0,\; \overline{\rho}_{i\tau}=0,\;
\overline{\nu}_{j\tau}=0,\; \overline{\rho}_{j\tau}>0,
\label{eq:symmetric_binding}
\end{equation}
which corresponds to a \emph{symmetric binding} pattern.

Under this condition, additional FRP is provided via \emph{unit dispatch transfer}: for FRU, capacity-bound units reduce output while ramp-limited units increase output, with the latter accommodating the additional FRU; for FRD, ramp-limited units reduce output while capacity-bound units increase output, with the ramp-limited units accommodating the additional FRD. When the units reducing output have lower generation costs than those increasing output, this redispatch raises total system cost, reflecting the flexibility constraint imposed by FRP and producing a suboptimal generator mix relative to FRP-unconstrained economic dispatch.

This phenomenon is illustrated using a two-unit system over a two-interval window $\mathscr{H}_t = \{t, t+1\}$, with $G_1$ (low-cost, low-emission) and $G_2$ (high-cost, high-emission) jointly provide FRP.

\begin{figure}[ht]
\centering
\includegraphics[width=0.48\textwidth]{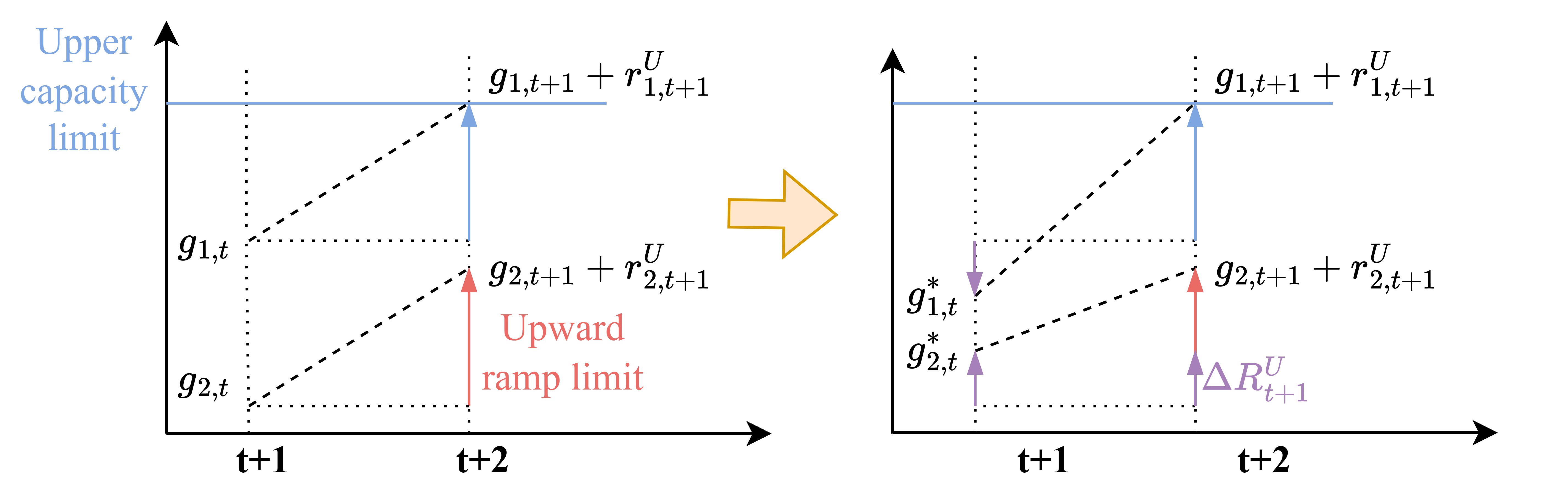}
\caption{$G_1$ binds at its upper capacity and $G_2$ at its ramp-up limit; additional FRU requirement induces a dispatch transfer from $G_1$ to $G_2$.}
\label{fig:FRU_boundary}
\end{figure}

\begin{figure}[ht]
\centering
\includegraphics[width=0.48\textwidth]{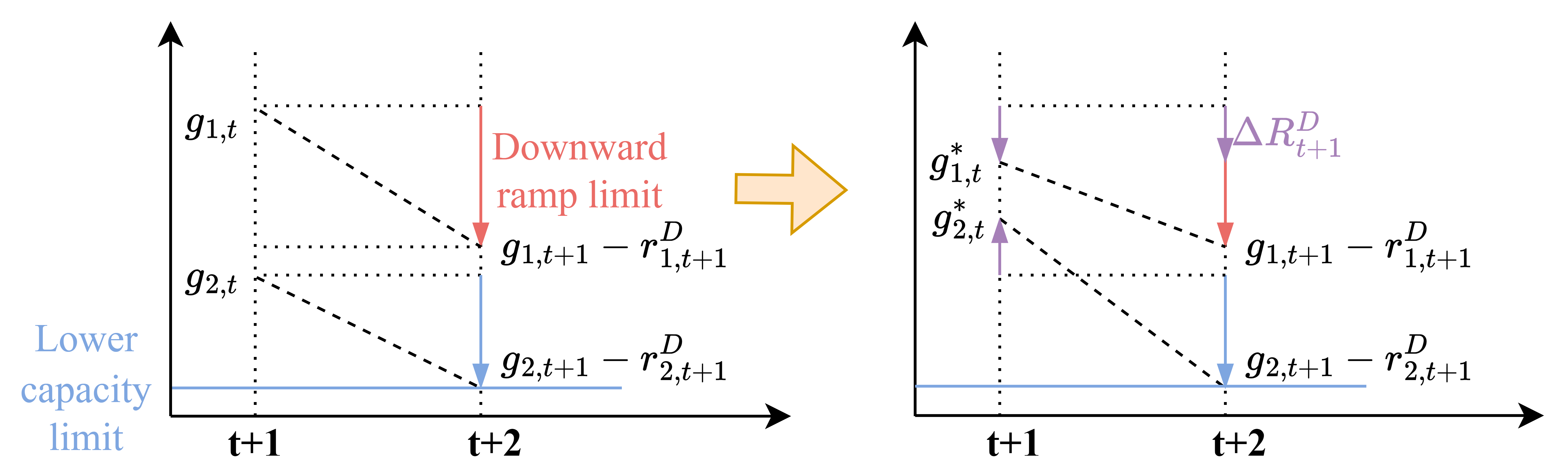}
\caption{$G_1$ binds on ramp-down limit and $G_2$ on lower capacity; additional FRD requirement induces a dispatch transfer from $G_1$ to $G_2$.}
\label{fig:FRD_boundary}
\end{figure}

For FRU, $\phi_{t+1}^U>0$ arises when $G_1$ reaches upper capacity and $G_2$  ramp-up limit in the advisory interval $t+1$, triggering a transfer where $G_2$ increases output while $G_1$ decreases in binding interval $t$ (Fig.~\ref{fig:FRU_boundary}). Conversely, $\phi_{t+1}^D>0$ occurs when $G_1$ binds on ramp-down and $G_2$ on lower capacity (Fig.~\ref{fig:FRD_boundary}).

These results show that high FRP requirements can restrict system flexibility, forcing the dispatch of units with higher generation costs and emissions.  Section~\ref{sec:Cap_Mode} introduces a \emph{regulated forecast-based dispatch} approach to mitigate this effect.

\section{Regulated Forecast-based Dispatch Approach}\label{sec:Cap_Mode}

The previous model follows CAISO practice and is referred to as the FBD mode, where VERs participate in the market according to their forecast values (see Equation~\ref{eq:balance}). To improve flexibility under high VER penetration, we introduce a \emph{regulated forecast-based dispatch} approach. The power balance in this mode is:

\begin{equation}
\lambda_{\tau}: \sum_{i \in N_G} g_{i\tau} 
= \sum_{l \in N_D} (\hat{d}_{l\tau|t} - \delta d_{l\tau})_ - \hat{V}_{\tau|t},
\end{equation}
where \(\hat{V}_{\tau|t}\) denotes the total VER generation at interval \(\tau\) based on the capped values.

 Consider a rolling window \(\mathscr{H}_t = \{t, \dots, t+W-1\}\). The treatment of VERs differs for the binding interval (\(\tau = t\)) and the advisory intervals (\(\tau = t+1, \dots, t+W-1\)). 

\subsection{ Advisory Intervals (\(\tau = t+1, \dots, t+W-1\))}

For these intervals, the VER forecasts are not perfectly reliable, so we apply a \emph{mathematical cap} to reduce potential over-forecasting:
\begin{equation}
\hat{v}_{k\tau|t}^{\mathrm{cap}} = \hat{v}_{k\tau|t} - \Delta_{k\tau}^{\mathrm{cap}}, 
\quad \Delta_{k\tau}^{\mathrm{cap}} \ge 0, 
\quad \forall k \in N_v, \ v \in \{w,s\}.
\end{equation}

The total capped generation is then
\begin{equation}\label{eq:advisory_cap}
\hat{V}_{\tau|t} = \sum_{v \in \{w,s\}} \sum_{k \in N_v} \hat{v}_{k\tau|t}^{\mathrm{cap}}, 
\quad \tau = t+1, \dots, t+W-1.
\end{equation}

Here, the cap does not directly correspond to the actual physical generation but helps reduce FRP requirements in the advisory horizon.

\subsection{ Binding Interval (\(\tau = t\))}

For the binding interval, the forecast is assumed to be perfect (\(\hat{v}_{k t|t}\) represents realized generation for interval $t$). The total VER generation is limited by the minimum between the realized forecast and the capped forecast from the previous rolling window \(\mathscr{H}_{t-1}\):
\begin{equation}\label{eq:capped_forecast}
\hat{V}_{t|t} = \min \Bigg(
\sum_{v \in \{w,s\}} \sum_{k \in N_v} \hat{v}_{k t|t}, 
\sum_{v \in \{w,s\}} \sum_{k \in N_v} \hat{v}_{k t|t-1}^{\mathrm{cap}}
\Bigg).
\end{equation}

\begin{figure}[ht]
\centering
\includegraphics[width=0.45\textwidth]{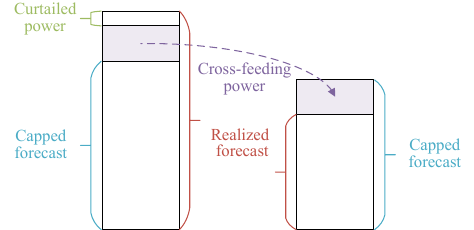}
\caption{Cross-feeding among two VER units}
\label{fig:cross-feeding}
\end{figure}

This total-cap formulation implicitly enables \emph{cross-feeding} among VER units, as illustrated in Fig.~\ref{fig:cross-feeding}: a shortfall in one unit can be offset by another exceeding its individual cap, enabling renewables to mutually mitigate variability.

\subsection{FRP Requirements}\label{sec:RFBD_frp}

The aggregated VER uncertainty for interval $t+1$ under RFBD is
\begin{equation}
\Delta V_{t+1} = \hat{V}_{t+1|t+1} - \hat{V}_{t+1|t}.
\end{equation}

For simplicity, we assume perfect load forecasts and determine the FRP requirements based on the histogram quantiles of the net load, a basic method used by CAISO \cite{CAISO_Appendices_Market_Operations_V62}. Specifically, a frequency histogram of the net load changes is constructed, and the 2.5\% and 97.5\% quantiles are taken as the downward and upward FRP requirements, respectively.

Accordingly, the net load uncertainty is defined separately for the two dispatch modes:

\begin{equation}
\Delta L_{t+1}^{\mathrm{FBD}} = - \Delta W_{t+1} - \Delta S_{t+1},
\end{equation}
where $\Delta W_{t+1}$ and $\Delta S_{t+1}$ represent the total forecast errors of wind and solar generation, respectively.

\begin{equation}
\Delta L_{t+1}^{\mathrm{RFBD}} = - \Delta V_{t+1}.
\end{equation}

\begin{property}\label{theorem1}
Under RFBD, the FRD requirement is eliminated.
\end{property}

\begin{proof}
From the definitions of $\hat{V}_{t+1|t}$ (Equation~\ref{eq:advisory_cap}) and $\hat{V}_{t+1|t+1}$ (Equation~\ref{eq:capped_forecast}), it follows that
\begin{equation}
\Delta V_{t+1} = \hat{V}_{t+1|t+1} - \hat{V}_{t+1|t} \leq 0,
\end{equation}
which implies
\begin{equation}
\Delta L_{t+1}^{\mathrm{RFBD}} = - \Delta V_{t+1} \geq 0.
\end{equation}

Since the FRD requirement is determined by the negative tail of the distribution of $\Delta L_{t+1}^{\mathrm{RFBD}}$, and this quantity is always nonnegative, no FRD is needed under RFBD.
\end{proof}

\begin{property}\label{theorem2}
Under RFBD, the FRU requirement is reduced compared to FBD, with the reduction amount equal to the total capping values.
\end{property}

\begin{proof}

Assume that
\[
\sum_{v \in \{w,s\}} \sum_{k \in N_v} \hat{v}_{k,t|t} \le \sum_{v \in \{w,s\}} \sum_{k \in N_v} \hat{v}_{k,t|t-1}^{\mathrm{cap}}.
\]

Substituting \(\hat{v}_{k, t+1|t}^{\mathrm{cap}} = \hat{v}_{k, t+1|t} - \Delta_{k,t+1}^{\mathrm{cap}}\), we obtain
\begin{equation}
\Delta L_{t+1}^{\mathrm{RFBD}} 
= \sum_{v \in \{w,s\}} \sum_{k \in N_v} \big( \hat{v}_{k, t+1|t} - \Delta_{k,t+1}^{\mathrm{cap}} - \hat{v}_{k, t+1|t+1} \big).
\end{equation}

In contrast, under the forecast-based dispatch mode,
\begin{equation}
\Delta L_{t+1}^{\mathrm{FBD}} 
= \sum_{v \in \{w,s\}} \sum_{k \in N_v} \big( \hat{v}_{k, t+1|t} - \hat{v}_{k, t+1|t+1} \big).
\end{equation}

Therefore, the difference between the two formulations is
\begin{equation}
\Delta L_{t+1}^{\mathrm{FBD}} - \Delta L_{t+1}^{\mathrm{RFBD}} 
= \sum_{v \in \{w,s\}} \sum_{k \in N_v} \Delta_{k,t+1}^{\mathrm{cap}}.
\end{equation}

Since the FRU requirement is determined by the right tail (97.5\% quantile) of the distribution of net load uncertainty, the RFBD reduces the required FRU by exactly the amount of total capping \(\sum_{v \in \{w,s\}} \sum_{k \in N_v} \Delta_{k\tau}^{\mathrm{cap}}\).
\end{proof}

\section{Case Studies}

\subsection{Settings}

Two conventional generators and a single load are considered, with their settings summarized in Table~\ref{tab:gen_settings}.

\begin{table}[h]
\centering
\caption{Generator and load shedding settings}
\label{tab:gen_settings}
\resizebox{0.45\textwidth}{!}{%
\begin{tabular}{c|c|c|c|c|c}
\hline
Unit & $C_i^g$ (\$/MWh) & $[\underline{G}_i, \overline{G}_i]$ (MW) & $[\underline{r}_i, \overline{r}_i]$ (MW) & $C^L$ & $e_i$ (tCO$_2$/MWh) \\
\hline
$G_1$ & 20 & $[0,100]$ & $[-15,15]$ & -- & 0.214 \\
$G_2$ & 50 & $[0,500]$ & $[-50,50]$ & -- & 0.428\protect\footnotemark \\
$D$ & -- & -- & -- & 200 & -- \\
\hline
\end{tabular}%
}
\end{table}

\footnotetext{{Following the GHG RP default emission factor for unspecified sources~\cite{CAISO2024GHGPresentation}}}

Two VERs are modeled with i.i.d. Gaussian forecast errors:
\begin{equation}
\hat{v}_{k,\tau+1|\tau+1} \sim \mathcal{N}(\hat{v}_{k,\tau+1|\tau}, (0.1 \cdot \hat{v}_{k,\tau+1|\tau})^2), \quad k=1,2.
\end{equation}

A two-stage cascaded rolling-window dispatch is applied, with windows 
$\mathscr{H}_t = \{t, t+1\}$ and $\mathscr{H}_{t+1} = \{t+1, t+2\}$, and the detailed settings are listed in Table~\ref{tab:rolling_window_settings}. 
In $\mathscr{H}_{t+1}$, the realized forecast for interval \(t+1\) (i.e., $\hat{v}_{k,t+1|t+1}$) may deviate from the advisory forecast $\hat{v}_{k,t+1|t}$ in $\mathscr{H}_{t}$. 
Such deviations are characterized using Monte Carlo sampling.

\begin{table}[h]
\centering
\caption{Rolling-window dispatch settings}
\label{tab:rolling_window_settings}
\begin{tabular}{c|c|c|c|c}
\hline
Window & Interval & $\hat{d}$ (MW) & $\hat{v}_1$ (MW) & $\hat{v}_2$ (MW) \\
\hline
\multirow{2}{*}{$\mathscr{H}_t$} & $t$ & 100 & 20 & 20 \\
 & $t+1$ & 85 & 20 & 20 \\
\hline
\multirow{2}{*}{$\mathscr{H}_{t+1}$} & $t+1$ & 85 & $\mathcal{N}(20,4)$ & $\mathcal{N}(20,4)$ \\
 & $t+2$ & 85 &  20 & 20 \\
\hline
\end{tabular}
\end{table}

Four dispatch modes are compared:  
\begin{enumerate}
    \item \textbf{FBD:} Forecast-based dispatch without capping;  
    \item \textbf{Cap-$\Delta=0$:} FRD is eliminated while FRU remains unchanged;  
    \item \textbf{Cap-$\Delta=1$:} 1~MW curtailment per VER unit (5\% cap);  
    \item \textbf{Cap-$\Delta=2$:} 2~MW curtailment per VER unit (10\% cap).  
\end{enumerate}
  
For FBD, the initial conditions of $\mathscr{H}_{t}$ are set as $g_{1,t-1}=60$~MW and $g_{2,t-1}=0$~MW.  
A uniform capping is applied across all intervals; hence, $g_{1,t-1}$ increases to $62$~MW for Cap-$\Delta=1$ and to $64$~MW for Cap-$\Delta=2$, reflecting the previous-window effect.  
The optimal dispatch results $g_{1,t}$ and $g_{2,t}$ obtained from $\mathscr{H}_t$ are subsequently used as the initial conditions for $\mathscr{H}_{t+1}$.

\subsection{FRP Requirements}

A total of 1000 random samples are generated for $\hat{v}_{1,t+1|t+1}$ and $\hat{v}_{2,t+1|t+1}$ to represent forecast uncertainty.  
The FRP requirements for the advisory interval $t+1$ are then determined from the 2.5\% and 97.5\% quantiles of the net-load uncertainty histogram, constructed with a bin width of 0.5~MW, as illustrated in Fig.~\ref{fig:histogram}.  
The resulting FRU and FRD requirements are summarized in Table~\ref{tab:frp_results}.  

{As shown in Table~\ref{tab:frp_results}, FRD requirements are eliminated under RFBD, whereas FRU requirements are reduced in proportion to the total curtailed VER output ($2 \times \Delta$~MW for $\Delta$~MW curtailment per unit), validating Property~\ref{theorem1} and \ref{theorem2}.

\begin{figure}[ht]
\centering
\includegraphics[width=0.5\textwidth]{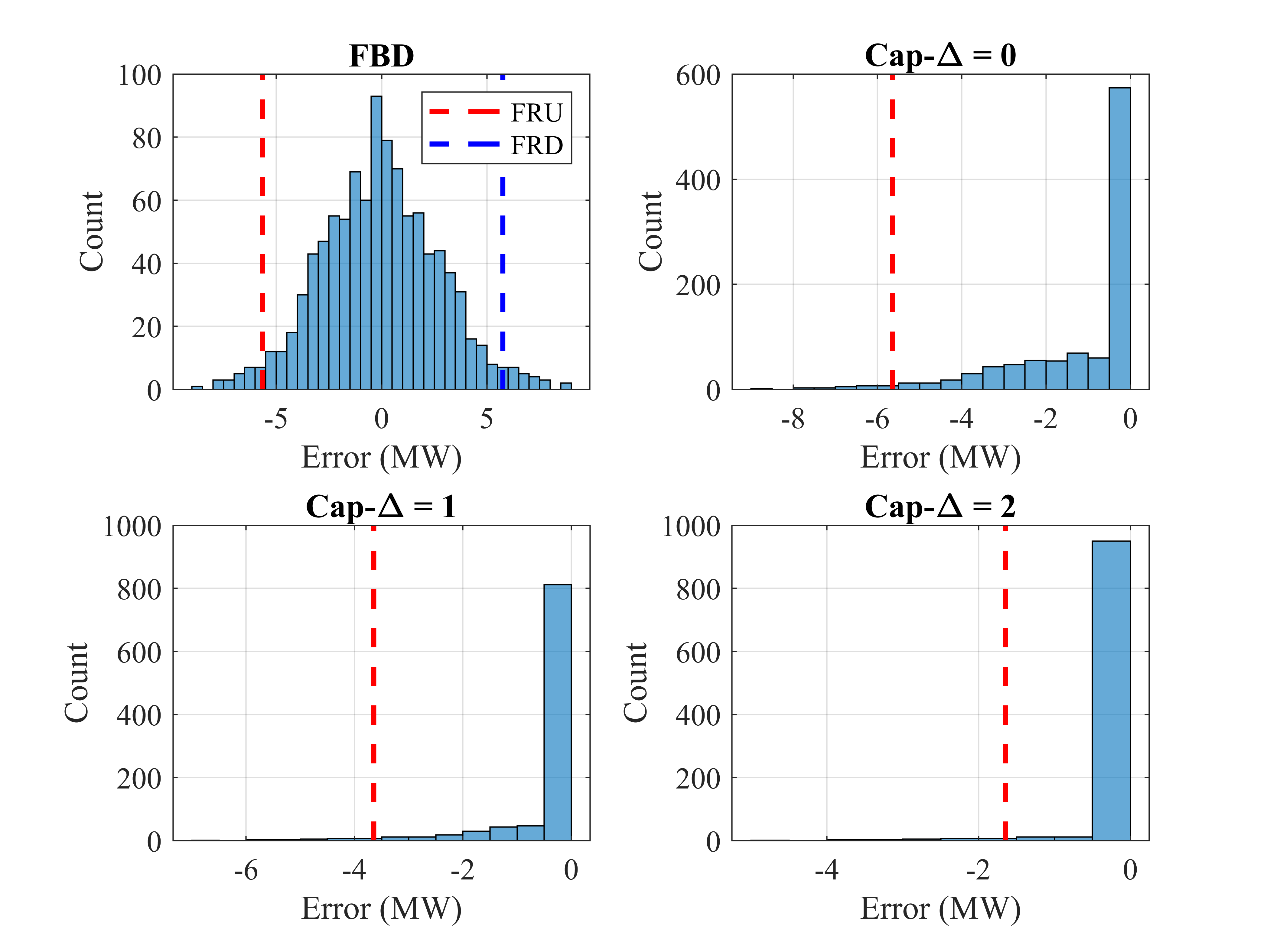}
\caption{Histogram of net load uncertainty for interval $t+1$}
\label{fig:histogram}
\end{figure}

\begin{table}[h]
\centering
\renewcommand{\arraystretch}{1.1}
\caption{FRP requirements at $t+1$ (MW)}
\label{tab:frp_results}
\begin{tabular}{c|cccc}
\hline
 & FBD & Cap-$\Delta=0$ & Cap-$\Delta=1$ & Cap-$\Delta=2$ \\
\hline
$R_{t+1}^U$ & 5.6451 & 5.6451 & 3.6451 & 1.6451 \\
$R_{t+1}^D$ & 5.7503 & 0 & 0 & 0 \\
\hline
\end{tabular}
\end{table}

\subsection{Results}

\begin{table}[h]
\centering
\renewcommand{\arraystretch}{1.2}
\caption{Optimal dispatch results for binding interval $t$, cost and carbon emissions for binding intervals $t$ and $t+1$}
\label{tab:combined_results}
\begin{tabular}{lcccc}
\toprule
 & FBD & Cap-$\Delta=0$ & Cap-$\Delta=1$ & Cap-$\Delta=2$ \\
\midrule
\multicolumn{5}{l}{\textbf{Binding interval $t$: dispatch (MW) and Lagrange multipliers}} \\
$g_{1,t}$ & 54.25 & 60 & 60 & 60 \\
$g_{2,t}$ & 5.75  & 0  & 0  & 0  \\
$\phi_{t+1}^U$ & 0  & 0  & 0  & 0  \\
$\phi_{t+1}^D$ & 30 & 0  & 0  & 0  \\
\midrule
\multicolumn{5}{l}{\textbf{Binding interval $t$: cost (\$) and emissions (tCO$_2$)}} \\
$C_t$ & 114.38 & 100.00 & 100.00 & 100.00 \\
$E_t$ & 1.173 & 1.070 & 1.070 & 1.070 \\
\midrule
\multicolumn{5}{l}{\textbf{Binding interval $t+1$: Monte Carlo average}} \\
$\bar{C}_{t+1}$ & 75.15 & 76.86 & 78.97 & 81.83 \\
$\bar{E}_{t+1}$ & 0.8041 & 0.8224 & 0.8450 & 0.8755 \\
\bottomrule
\end{tabular}
\end{table}

Table~\ref{tab:combined_results} summarizes the optimal dispatch results for the binding interval \(t\), as well as the cost and carbon emissions for the binding intervals \(t\) and \(t+1\). It can be observed that, under all cap values, RFBD yields \(\phi_{t+1}^D = \phi_{t+1}^U = 0\) in interval \(t\), thereby alleviating the flexibility constraints. Consequently, all Cap-\(\Delta\) cases share the same dispatch in interval \(t\). Figure~\ref{fig:cost_emission} shows the total cost and emissions across the two consecutive binding intervals. RFBD generally outperforms FBD, validating the effectiveness of RFBD in reducing operating cost.
Since FRP has zero bidding cost in CAISO, larger caps reduce the FRU requirement but do not lower the operating cost. 
Therefore, Cap-$\Delta = 0$ achieves the minimum cost.

\begin{figure}[ht]
\centering
\includegraphics[width=0.5\textwidth]{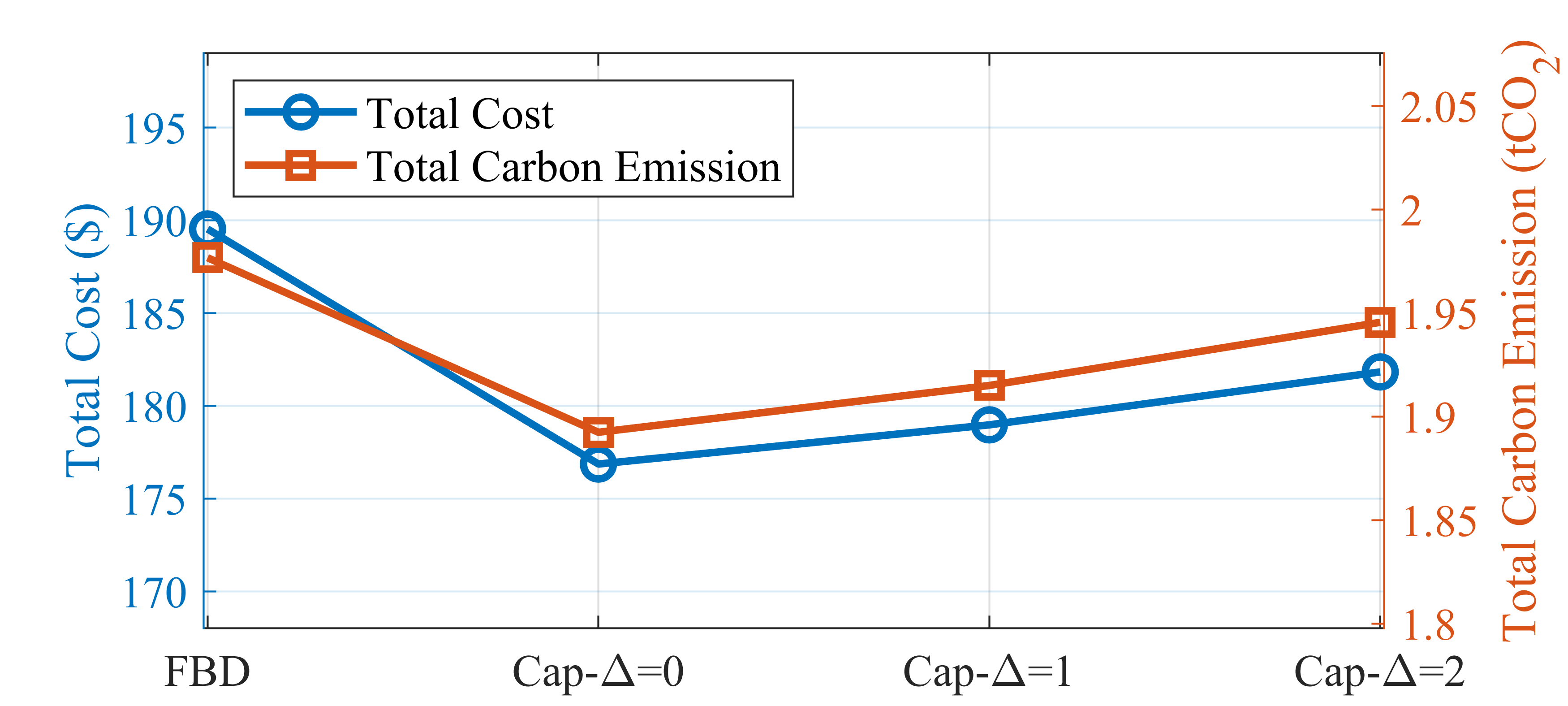}
\caption{Total cost and carbon emissions across two consecutive binding intervals.}
\label{fig:cost_emission}
\end{figure}

\section{Discussion}

The energy–FRP co-optimization model is formulated under the assumption that transmission congestion is neglected. In practice, network constraints may limit the feasibility of cross-feeding among renewable units, and the inclusion of unit commitment introduces additional operational considerations, such as start-up costs and minimum up times, which may affect optimal dispatch results. Therefore, extending the proposed framework to incorporate network constraints and unit commitment, as well as validating it on larger-scale networks, remains an important direction for future work.

In addition, the regulatory implications of deliberate pre-curtailment are important for practical relevance but require in-depth consultation with system operators. Accordingly, regulatory considerations are also left for future work.

\section{Conclusion}

This paper theoretically reveals that high FRP requirements under FBD constrain real-time flexibility and lead to suboptimal generator dispatch, whereas the proposed RFBD approach effectively reduces FRP requirements and operating cost. This approach provides a practical means to reduce uncertainty and enhance flexibility in renewable-rich power systems.

\bibliographystyle{IEEEtran}
\bibliography{reference}

\end{document}